\documentclass[%
 reprint,
 amsmath,amssymb,
 aps,
 pra,
 prb,
 rmp,
 prstab,
 prstper,
 floatfix,
 ]{revtex4-1}
\usepackage{amsmath}
\usepackage{amssymb}
\usepackage{units}
\usepackage{color}
\usepackage{graphicx}
\usepackage{dcolumn}
\usepackage{bm}
\usepackage{fixmath}
\usepackage{float}
\usepackage{amsmath}

\newsavebox{\savedimage}

\begin{document}

\title{Role of enviromental dynamic polarizability in static excited state properties of embedded molecular systems: Application to disordered fluorographene systems}
\author{Vladislav Sl\'ama$^{1}$}
\email{slama@karlov.mff.cuni.cz}
\author{Frank M\"uh$^{2}$}
\author{Thomas Renger$^{2}$}
\author{Tom\'a\v{s} Man\v{c}al$^{1}$}
\affiliation{$^{1}$Faculty of Mathematics and Physics, Charles University, Ke Karlovu 5, 121 16 Prague 2, Czech Republic, \\
$^{2}$Institut f\"{u}r Theoretische Physik, Johannes Kepler University Linz, Altenberger Str. 69, 4040 Linz, Austria}

\begin{abstract}
We present an extension of the standard polarizable quantum mechanic/molecular mechanics (QM/MM) approach for treating environmental effects on excited state properties of embedded systems. A quantum polarizable atom model is derived from a full quantum description of the environment using perturbation theory for the system-environment coupling. Our model provides a more general description of the environment, including frequency dependent atomic polarizabilities, without loss of computational simplicity of the QM/MM approach. The classical polarizable atom description used in polarizable QM/MM approach can be obtained as a limiting case of the present model. This enables us to pinpoint approximations and assumptions about the system inherent to classical polarizable atom description of the environment. We apply our method to fluorographene (FG) based systems to investigate effects of the FG sheet on excited state properties of impurities on the FG surface. Comparison to full quantum chemistry calculations and to the standard polarizable QM/MM approach on disordered fluorograhene clusters reveals the importance of frequency dependence of atomic polarizabilities for the proper treatment of the effects of the environment on both the transition energy shifts and the interaction energies.
\end{abstract}

\maketitle

\section{Introduction}
The effects of environment play a crucial role in photochemistry and photophysics of embedded molecular systems. In complex systems with many chromophores, environment influences excited state properties and dynamics in variety of ways. For instance, different environments of different sides of otherwise symmetric chromophore structures lead to an asymmetric electron transfer \cite{MuhRenRC2017}. Changes in environmental conformations lead to switching between different excitation energy pathways, creation of traps or even quenching of the excitation \cite{DuffyNPQ2013}. The environment usually accomplishes the feat by modulating the chromophore site energies and interaction energies between individual chromophores. Correct description of environmental effects on the chromophore excited state properties is crucial for understanding the function of many photochemical systems. 

In the past years, substantial effort has been devoted to development and improvement of models that account for the effects of environment in excited state quantum chemical calculations. These methods can be divided into quantum subsystem approaches (QM/QM), combined quantum mechanical and molecular mechanical methods (QM/MM) and polarizable continuum models (PCM). In QM/QM methods the whole system is divided into subsystems which are described with quantum chemistry methods \cite{GordonEFP2001,WesolFrDensEmb2015}. Within this approach, the heterogenity of the environment is well described, however, computational cost limits the number of subsystems which could be feasibly considered. In the PCM models \cite{CossiPCM2000} on the other hand, the solvent is represented as a polarizable continuum medium characterized by its macroscopic dielectric properties, whereas the solute, located in a molecular-shaped cavity inside the dielectric, is described at a full QM level. The PCM models are very popular nowadays due to their efficiency and accuracy in describing homogeneous environments, where no strong specific solvent-solute interactions are present. This, however, makes these methods unsuitable for many biological systems, where chromophores are embedded in protein environment, the heterogeneity of which is crucial for the system's function \cite{RengerFMO2006}. Hybrid QM/MM approach presents a good alternative way. In QM/MM methods, the system of interest is described quantum mechanically, whereas the environment is described by a classical force field. This reduces the computational cost considerably, compared to QM/QM methods, while keeping the atomic description of the environment. In the past decades, these methods were successfully applied to various biological systems. In order to account for the non-equilibrium polarizability of the environment, the polarizable MM forcefields were developed in form of the QM/MMpol \cite{Curutchet2009} type of methods. Mutual polarization of the chromophore system and the environment is in these methods solved until self consistency is reached. The description of the environment is classical here, assigning every atom a static (i.e. frequency independent) polarizability. Polarizable QM/MM approach was successfully applied to investigation of excited state properties of many biological systems such as LH2 \cite{SegattaPolMMLH22017}, FMO \cite{Curutchet2014} or small organic molecules in water solution \cite{Mennucci2009,Mennucci2015}. An example of a system where this approximation fails and the frequency dependent polarizabilities are required for a correct description of environmental effects is disordered fluorographene, which we study in this work.

Fluorographene (FG) was recently introduced as a promising medium for artificial light harvesting systems \cite{Slama2018}. It was proposed that graphene-like impurities, resulting from removal of fluorine atoms from the perfectly fluorinated FG, could serve as centers, quantum dots, for the excitation energy transfer. It was shown by quantum chemical calculations on small FG clusters that $\pi$-conjugated impurities exhibit transitions inside the FG energy gap with corresponding transition densities localized in the area of the impurity. This makes them treatable as individual molecules embedded in pure FG environment. FG based systems also enjoy many important properties of good artificial light harvester, such as photochemical and thermal stability and the possibility to tune the sensitivity to a broad range of wavelengths in solar spectrum. To further study the prospects of impurities in FG could as efficient light harvesters, structure based calculations are needed. Two important interactions that need to be described are impurity-impurity and impurity-environment coupling. Treatment of large FG systems with many impurities and investigation of effects of different impurities and their mutual configurations on efficiency of excitation energy transfer is not in reach of the full QC treatment of the system. Therefore a simpler but reliable approximate structure-based method is needed. Substantial covalent bonding between impurity and the FG and close distance between FG atoms makes the QM/MM treatment complicated and unsuitable for the FG based systems.

In this article we present a general QM/QM approach inspired by Ref. \cite{Renger2013} for the description of the environment effects on chromophore systems. We start from the properties of isolated chromophores in vacuum, which can be obtained using standard QC methods. Our method is based on perturbation treatment of the system-environment interaction resulting in quantum polarizable atom description of the environment. The quantum polarizable atom model provides a generalization of the standard classical polarizable atom model. The standard approximation of the environment atoms as harmonic oscillators, resulting in classical polarizable atom model, can be obtained as a limiting case of the presented approach. This makes it possible to pinpoint the approximations and assumptions about the system inherent to the classical polarizable atom model. The quantum polarizable atom model derived here, is then applied to calculation of environment induced site energy and transition dipole shifts and interaction energies between excited states of impurities in FG. Results from quantum polarizable atom approach are compared with full quantum chemical (QC) calculation, with PoisonTrEsp method which uses polarizable continuum description of the environment and with the classical polarizable atom model of the environment, on set of small FG clusters. Quantum polarizable atom model represents a general description of embedded systems and an improvement of the classical polarizable atom model. When applied to disordered fluorographene systems, it provides computationally very efficient and reliable alternative to full quantum chemical (QC) approach.

The paper is organized as follows: in Section \ref{sec:theory}, we outline perturbation theory of the environmental effects on transition energies and transition dipole moments of system embedded in a quantum polarizable medium. We develop a theory of environmental effects on the resonance coupling between two such systems and discuss classical polarizable atom model as a limit of the quantum model we develop here. In Section \ref{sec:discussion}, we discuss calculations of excited state properties of impurities in FG, which serve as validity demonstration of the quantum polarizable atom model. We sum up our conclusions in Section \ref{sec:conclusions}.

\section{Theory}
\label{sec:theory}
Let us start with considering a pair of chromophores, denoted $A$ and $B$, interacting with certain environmental building blocks $n$. There is a considerable freedom in the choice of the individual environmental building blocks. One may choose single atoms or small groups of atoms (like the present case of FG) or whole molecules (like amino-acids in the case of proteins), depending on the level of coarse graining that the material allows. In both the cases we mention, the environmental blocks have $\sigma$-bond structures, resulting in their higher excitation energies then those of the $\pi$-conjugated chromophores of the impurities in FG or pigments in photosynthetic complexes. The large energy gap between the system transitions and those of its environment makes the quantum perturbation approach suitable to account for the environmental effects on chromophore excited state properties in these systems.

Let us denote the localized states of individual parts of the system as $\left|a_{A}\right\rangle$ , $\left|b_{B}\right\rangle$  and $\left|\eta_{n}\right\rangle$  for the states where the chromophore $A$ is in the state $a$, chromophore $B$ in the state $b$ and $n$th environment building block in the state $\eta$, respectively. The respective energies are denoted as $\varepsilon_{a}^{\left(A\right)}$, $\varepsilon_{b}^{\left(B\right)}$ and $F_{\eta}^{(n)}$. The individual parts of the system are interacting through Coulomb interaction. The total Hamiltonian can be written as $H=H_{A}+H_{B}+\sum_{n}H_{n}+V$ where $H_{i},\;(i=A,B,n)$ are the Hamiltonian operators of non-interacting system parts and V is an interaction operator. The interaction operator can be separated into $V=V_{AB}+\sum_{n}V_{An}+\sum_{n}V_{Bn}+\frac{1}{2}\sum_{nn'}V_{nn'}$ where $V_{An}$ and $V_{Bn}$ correspond to Coulomb interaction of the chromophore with environment building block $n$ and $V_{nn'}$ to mutual interaction between environment building blocks $n$ and $n'$. We assume no molecular orbital overlaps between individual parts of the system. In the absence of electron exchange between individual parts, we can use the basis functions in form of Hartree product $\left|ab,\boldsymbol{\eta}\right\rangle =\left|a_{A}\right\rangle \left|b_{B}\right\rangle \prod_{n}\left|\eta_{n}\right\rangle$  to investigate the effects of the environment-chromophore interaction on the chromophore properties. The multiindex $\boldsymbol{\eta}=\left(\eta_{1},\eta_{2},\dots\eta_{n},\dots\right)$ is introduced to abbreviate the notation for electronic states of the environment. For representation purposes we assume only two level chromophores, with ground and excited electronic states, however, the approach is general and can be applied to chromophores with many excited states.

\subsection{Transition energy}

For investigating the effects of the environment on the transition energy we will assume only single chromophore interacting with its environment. Site energy shift $\Delta E_{a}=E_{a}-\varepsilon_{a}^{\left(A\right)}$, corresponds to the change from the transition energy $\varepsilon_{a}^{\left(A\right)}$ of the chromophore in vaccum to the transition energy $E_{a}$ of the chromophore embedded in the environment. The zero and the first order contributions of the perturbation expansion to the chromophore energy shift $\Delta E_{a}$, can be written as
\begin{widetext}
\begin{equation} \label{eq:EnergyShift_first-order}
\begin{split}
&\Delta E_{a}^{(0+1)}=\sum_{n}\left\langle a,\boldsymbol{0}\right|V_{An}\left|a,\boldsymbol{0}\right\rangle +\frac{1}{2}\sum_{n,n'}\left\langle a,\boldsymbol{0}\right|V_{nn'}\left|a,\boldsymbol{0}\right\rangle -\sum_{b}\sum_{\boldsymbol{\eta}}\frac{\left|\left\langle a,\boldsymbol{0}\right|V\left|b,\boldsymbol{\eta}\right\rangle \right|^{2}}{F_{0\eta}^{\left(n\right)}+\varepsilon_{ab}^{\left(A\right)}} \\ 
&=\sum_{n}\left\langle a_{A}\right|\left\langle 0_{n}\right|V_{An}\left|0_{n}\right\rangle \left|a_{A}\right\rangle +\frac{1}{2}\sum_{n,n'}\left\langle 0_{n}\right|\left\langle 0_{n'}\right|V_{nn'}\left|0_{n'}\right\rangle \left|0_{n}\right\rangle -\sum_{n,\eta}\sum_{n'\ne n,\eta'}\frac{\left|\left\langle 0_{n}\right|\left\langle 0_{n'}\right|V_{nn'}\left|\eta_{n}\right\rangle \left|\eta'_{n'}\right\rangle \right|^{2}}{\left(F_{0\eta}^{\left(n\right)}+F_{0\eta'}^{\left(n'\right)}\right)}\\
&-\sum_{n,\eta}\sum_{n'}\frac{\left|\left\langle 0_{n'}\right|\left\langle 0_{n}\right|V_{nn'}\left|\eta_{n}\right\rangle \left|0_{n'}\right\rangle \right|^{2}}{F_{0\eta}^{\left(n\right)}}-\frac{\sum_{n}\left|\left\langle a_{A}\right|\left\langle 0_{n}\right|V_{An}\left|0_{n}\right\rangle \left|b_{A}\right\rangle \right|^{2}}{\varepsilon_{ab}^{\left(A\right)}} - \sum_{n,\eta}\frac{\left|\left\langle a_{A}\right|\left\langle 0_{n}\right|V_{An}\left|\eta_{n}\right\rangle \left|a_{A}\right\rangle \right|^{2}}{F_{0\eta}^{\left(n\right)}}\\
&-\sum_{n,\eta}\frac{\left|\left\langle a_{A}\right|\left\langle 0_{n}\right|V\left|\eta_{n}\right\rangle \left|b_{A}\right\rangle \right|^{2}}{F_{0\eta}^{\left(n\right)}+\varepsilon_{ab}^{\left(A\right)}}-2\sum_{n,\eta}\sum_{n'}\frac{\left\langle a_{A}\right|\left\langle 0_{n}\right|V_{An}\left|\eta_{n}\right\rangle \left|a_{A}\right\rangle \left\langle 0_{n'}\right|\left\langle \eta_{n}\right|V_{nn'}\left|0_{n}\right\rangle \left|0_{n'}\right\rangle }{F_{0\eta}^{\left(n\right)}},
\end{split}
\end{equation}
\end{widetext}
where $F_{0\eta}^{\left(n\right)}=F_{\eta}^{\left(n\right)}-F_{0}^{\left(n\right)}$ and $\varepsilon_{ab}^{\left(A\right)}=\varepsilon_{b}^{\left(A\right)}-\varepsilon_{a}^{\left(A\right)}$ correspond to the excitation energy of an isolated environmental building block $n$ and the chromophore $A$, respectively. The first term in Eq. (\ref{eq:EnergyShift_first-order}) corresponds to Coulomb interaction between the chromophore state $a$ charge density and the environmental ground state charges. This can be approximated by Coulomb coupling between atomic partial charges \cite{Madjet2006}. The electrostatic interaction energy reads as
\begin{equation} \label{eq:ElstatInter-def}
\begin{split}
E_{aa}^{ele}&=\sum_{n}\left\langle a_{A}\right|\left\langle 0_{n}\right|V_{An}\left|0_{n}\right\rangle \left|a_{A}\right\rangle\\
&=\sum_{n}\sum_{J\in n}\phi_{a}^{\left(A\right)}\left(\boldsymbol{R}_{J}\right)q_{J}^{\left(n\right)}\left(0,0\right) \\
&=\sum_{n}\sum_{J\in n}\sum_{I\in A}\frac{q_{I}^{\left(A\right)}(a,a)}{\left|\boldsymbol{R}_{J}-\boldsymbol{R}_{I}\right|}q_{J}^{\left(n\right)}\left(0,0\right),
\end{split}
\end{equation}
where $\phi_{a}^{\left(A\right)}\left(\boldsymbol{R}_{J}\right)$ is the electrostatic potential generated by chromophore charge density at electronic state $a$ and $q_{J}^{\left(n\right)}\left(0,0\right)$ is atomic charge on atom $J$ at position $\boldsymbol{R}_{J}$. The total electrostatic potential of the chromophore at the state $a$ is approximated by the potential of a set of point atomic charges $q_{I}^{\left(A\right)}(a,a)$ as $\phi_{a}^{\left(A\right)}\left(\boldsymbol{R}\right)=\sum_{I\in A}q_{I}^{\left(A\right)}(a,a)/\left|\boldsymbol{R}-\boldsymbol{R}_{I}\right|$. The second, third and fourth terms in Eq. (\ref{eq:ElstatInter-def}) correspond to interaction between environmental building blocks. This interaction is independent of the chromophore state and therefore it does not contribute to the transition energy shift. 

The term $\sum_{n}\left|\left\langle a_{A}\right|\left\langle 0_{n}\right|V_{An}\left|0_{n}\right\rangle \left|b_{A}\right\rangle \right|^{2}$ in Eq. (\ref{eq:EnergyShift_first-order}) corresponds to Coulomb interaction of the chromophore transition density (charges) with the environmental ground state charges
\begin{equation} \label{eq:ElstatInter-transition-def}
\begin{split}
E_{ab}^{ele}&=\sum_{n}\left\langle a_{A}\right|\left\langle 0_{n}\right|V_{An}\left|0_{n}\right\rangle \left|b_{A}\right\rangle \\ 
&=\sum_{n}\sum_{J\in n}\sum_{I\in A}\frac{q_{I}^{\left(A\right)}(a,b)}{\left|\boldsymbol{R}_{J}-\boldsymbol{R}_{I}\right|}q_{J}^{\left(n\right)}\left(0,0\right).
\end{split}
\end{equation}
Applying multipole expansion to the Coulomb interaction, we find that the term $\left\langle a_{A}\right|\left\langle 0_{n}\right|V_{An}\left|\eta_{n}\right\rangle \left|a_{A}\right\rangle$, of Eq. (\ref{eq:EnergyShift_first-order}), can be interpreted as a chromophore charge density interaction with the transition dipole $\boldsymbol{\mu}_{\eta}^{\left(n\right)}$ of the environmental building block $n$. We obtain
\begin{equation} \label{eq:EnergyShift-PolarStaticEnergy}
\begin{split}
\left\langle A_{a}\right|\left\langle 0_{n}\right|V_{An}\left|\eta_{n}\right\rangle \left|A_{a}\right\rangle =-\boldsymbol{\mathcal{E}}_{aa}^{\left(A\right)}\left(\boldsymbol{R}_{n}\right)\boldsymbol{\mu}_{\eta}^{\left(n\right)} \\
=-\sum_{I\in A}\frac{q_{I}^{\left(A\right)}(a,a)\left(\boldsymbol{R}_{n}-\boldsymbol{R}_{I}\right)}{\left|\boldsymbol{R}_{n}-\boldsymbol{R}_{I}\right|^{3}}\vec{\mu}_{\eta}^{\left(n\right)},
\end{split}
\end{equation}
where $\boldsymbol{\mathcal{E}}_{aa}^{\left(A\right)}\left(\boldsymbol{R}_{n}\right)$ is the electric field generated by the chromophore state $a$ charge density at the position of the center of the environmental building block $\boldsymbol{R}_{n}$. Inserting Eq. (\ref{eq:EnergyShift-PolarStaticEnergy}) back into the perturbation expansion, Eq. (\ref{eq:EnergyShift_first-order}), yields a contribution to the energy shift, which can be expressed in terms of the environment building block polarizations as
\begin{equation} \label{eq:EnergyShift-PolarStaticEnergy2}
\begin{split}
E_{aa,aa}^{pol,1}&\left(\alpha\left(0\right)\right) = -2\sum_{n,\eta}\frac{\left|\left\langle a_{A}\right|\left\langle 0_{n}\right|V_{An}\left|\eta_{n}\right\rangle \left|a_{A}\right\rangle \right|^{2}}{F_{0\eta}^{\left(n\right)}} \\
&=\sum_{I\in A}\sum_{J\in A}\sum_{n}q_{I}^{\left(A\right)}(a,a)q_{J}^{\left(A\right)}(a,a) \\
&\times\frac{\left(\boldsymbol{R}_{I}-\boldsymbol{R}_{n}\right)\overleftrightarrow{\alpha}^{\left(n\right)}\left(0\right)\left(\boldsymbol{R}_{n}-\boldsymbol{R}_{J}\right)}{\left|\boldsymbol{R}_{n}-\boldsymbol{R}_{I}\right|^{3}\left|\boldsymbol{R}_{n}-\boldsymbol{R}_{J}\right|^{3}}.
\end{split}
\end{equation}
Here, the polarizability tensor $\overleftrightarrow{\alpha}^{\left(n\right)}$ of the $n$th environment building block has been introduced. For the elements of the polarizability tensor $\alpha_{ij}^{\left(n\right)}$, we generalized the standard definition, found e.g. in Ref. \cite{Atkins2005}, to include dependency on transition energy E as
\begin{equation} \label{eq:Polarizability-def}
\alpha_{ij}^{\left(n\right)}\left(E\right)=2\sum_{\eta}\frac{\left(\boldsymbol{\mu}_{\eta}^{\left(n\right)}\right)_{i}\left(\boldsymbol{\mu}_{\eta}^{\left(n\right)}\right)_{j}}{F_{0\eta}^{\left(n\right)}-E}.
\end{equation}
The polarizability at zero frequency, $\overleftrightarrow{\alpha}^{\left(n\right)}\left(0\right)$, corresponds to the static polarizability of the environment building block, and  $\overleftrightarrow{\alpha}_{dynamic}^{\left(n\right)}\left(E\right)=\frac{1}{2}\left(\overleftrightarrow{\alpha}^{\left(n\right)}\left(E\right)+\overleftrightarrow{\alpha}^{\left(n\right)}\left(-E\right)\right)$ corresponds to the dynamic polarizability. Elements of $\overleftrightarrow{\alpha}_{dynamic}^{\left(n\right)}\left(E\right)$ are given as \cite{Atkins2005}
\begin{equation} \label{eq:DynamicPolar-def}
\left(\overleftrightarrow{\alpha}_{dynamic}^{\left(n\right)}\left(E\right)\right)_{ij}=2\sum_{\eta}\frac{F_{0\eta}^{\left(n\right)}\left(\boldsymbol{\mu}_{\eta}^{\left(n\right)}\right)_{i}\left(\boldsymbol{\mu}_{\eta}^{\left(n\right)}\right)_{j}}{\left(F_{0\eta}^{\left(n\right)}\right)^{2}-E^{2}}.
\end{equation}
The interpretation of Eq. (\ref{eq:EnergyShift-PolarStaticEnergy2}) is such that the chromophore $A$ excited (ground) state charges $q_{I}^{\left(A\right)}(a,a)$ produce an electric field $\boldsymbol{\mathcal{E}}_{aa}^{\left(A\right)}\left(R_{n}\right)$ at the location of environmental building block $n$. This field induces a dipole moment $\boldsymbol{\mu}_{aa}^{\left(n\right)ind,1}\left(\alpha\left(E\right)\right)=\overleftrightarrow{\alpha}^{\left(n\right)}\left(E\right)\boldsymbol{\mathcal{E}}_{aa}^{\left(A\right)}\left(R_{n}\right)$ which then interacts back with the charges on the same chromophore. The same interpretation is true also for the term $\sum_{n,\eta}\left|\left\langle a_{A}\right|\left\langle 0_{n}\right|V\left|\eta_{n}\right\rangle \left|b_{A}\right\rangle \right|^{2}/\left(F_{0\eta}^{\left(n\right)}+\left(\varepsilon_{ab}^{\left(A\right)}\right)\right)$, however, this time the dipole is induced by the electric field from the transition density (charges) assuming atomic polarizability $\overleftrightarrow{\alpha}^{\left(n\right)}\left(-\varepsilon_{ab}^{\left(A\right)}\right)$. The last term, denoted as $E_{aa}^{env-pol1}\left(\alpha\left(0\right)\right)$, in the first order energy perturbation expansion, Eq. (\ref{eq:EnergyShift_first-order}), corresponds to interaction of the chromophore state $a$ charge distribution with dipoles on the environmental blocks, induced by the ground state charges of the rest of the environment. This interaction can also be included into the total electrostatic interaction of the chromophore with its environment, which would then involve interaction of the chromophore with the static charges of the environmnet and with the dipoles induced by the environment itself.

For the purpose of this paper, we define $\boldsymbol{\mu}_{ab}^{\left(n\right)ind,i}\left(\alpha\left(E\right)\right)$ as the dipole on the environmental building blocks $n$ induced by the induced dipoles $\boldsymbol{\mu}_{ab}^{\left(n'\right)ind,i-1}\left(\alpha\left(E\right)\right)$ residing on all other environmental building blocks. The first dipole in the recurrent relation is induced by the chromophore charge density $\boldsymbol{\mu}_{ab}^{\left(n\right)ind,1}\left(\alpha\left(E\right)\right)=\overleftrightarrow{\alpha}^{\left(n\right)}\left(E\right)\boldsymbol{\mathcal{E}}_{ab}^{\left(A\right)}\left(R_{n}\right)$. We also define the energy $E_{ab,cd}^{pol,i}\left(\alpha\left(E\right)\right)=-\sum_{n}\mathcal{\vec{E}}_{ab}^{\left(A\right)}\left(R_{n}\right)\vec{\mu}_{ab}^{\left(n\right)ind,i}\left(\alpha\left(E\right)\right)$ of the interaction between the induced dipole $\boldsymbol{\mu}_{cd}^{\left(n\right)ind,i}\left(\alpha\left(E\right)\right)$ and the chromophore atomic charges $q_{I}^{\left(A\right)}(a,b)$.

Using the definitions above, the first order transition energy shift $\Delta E_{g\rightarrow e}^{\left(1\right)}=\Delta E_{e}^{(1)}-\Delta E_{g}^{(1)}$ can be written as
\begin{widetext}
\begin{equation} \label{eq:EnergyShift-FirstOrder}
\begin{split}
\Delta E_{g\rightarrow e}^{\left(1\right)}=\left(E_{ee}^{ele}-E_{gg}^{ele}\right)+\left[E_{ee}^{env-pol1}\left(\alpha\left(0\right)\right)-E_{gg}^{env-pol1}\left(\alpha\left(0\right)\right)\right]+\frac{2\left(E_{ab}^{ele}\right)^{2}}{\varepsilon_{ge}^{\left(A\right)}}\\
+\frac{1}{2}\left[E_{ee,ee}^{pol,1}\left(\alpha\left(0\right)\right)-E_{gg,gg}^{pol,1}\left(\alpha\left(0\right)\right)\right]
+\frac{1}{2}\left[E_{ge,ge}^{pol,1}\left(\alpha\left(\varepsilon_{ge}^{\left(A\right)}\right)\right)-E_{ge,ge}^{pol,1}\left(\alpha\left(-\varepsilon_{ge}^{\left(A\right)}\right)\right)\right].
\end{split}
\end{equation}
\end{widetext}
%
The first two terms in Eq. (\ref{eq:EnergyShift-FirstOrder}) correspond to a difference in electrostatic interaction between the chromophore and its environment for the chromophore in the excited and ground states. The third term represents the scaled electrostatic interaction of the transition density of the chromophore with its environment. The term $\left[E_{ee,ee}^{pol,1}\left(\alpha\left(0\right)\right)-E_{gg,gg}^{pol,1}\left(\alpha\left(0\right)\right)\right]$, in Eq. (\ref{eq:EnergyShift-FirstOrder}), can be interpreted as a solvation energy difference for the chromophore in the excited- and the ground states. The last term of Eq. (\ref{eq:EnergyShift-FirstOrder}) corresponds to the difference of solvation energies for chromophore transition density with two different environmental polarizabilities $\alpha\left(\varepsilon_{ab}\right)$ and $\alpha\left(-\varepsilon_{ab}\right)$.
The second order contribution to the chromophore excitation energy shift is obtained from the second order perturbation expansion following the same approach as for the first order. For the defects with large excitation energies compared to the electrostatic interaction of the defect with its environment, the second order correction to the transition energy shift reads as
\begin{widetext}
\begin{gather} 
\Delta E_{g\rightarrow e}^{\left(2\right)}\approx\frac{1}{2}\left[E_{ee,ee}^{pol,2}\left(\alpha\left(0\right)\right)-E_{gg,gg}^{pol,2}\left(\alpha\left(0\right)\right)\right]+\left[E_{ee}^{env-pol,2}\left(\alpha\left(0\right)\right)-E_{gg}^{env-pol,2}\left(\alpha\left(0\right)\right)\right]  \nonumber \\
+\frac{1}{4}\left[E_{ge,ge}^{pol,2}\left(\alpha\left(\varepsilon_{ge}^{\left(A\right)}\right)\right)
+E_{ge,ge}^{pol2}\left(\alpha\left(0\right),\alpha\left(\varepsilon_{ge}^{\left(A\right)}\right)\right)\right]
-\frac{1}{4}\left[E_{ge,ge}^{pol,2}\left(\alpha\left(-\varepsilon_{ge}^{\left(A\right)}\right)\right)+E_{ge,ge}^{pol2}\left(\alpha\left(0\right),\alpha\left(-\varepsilon_{ge}^{\left(A\right)}\right)\right)\right], \label{eq:EnergyShift-SecondOrder}
\end{gather}
\end{widetext}
where $E_{ge,ge}^{pol2}\left(\alpha\left(0\right),\alpha\left(\varepsilon_{ge}^{\left(A\right)}\right)\right)$ corresponds to the energy of interaction between the chromophore and the first order induced dipoles assuming the frequency dependent atomic polarizability $\alpha\left(\varepsilon_{ge}^{\left(A\right)}\right)$, and between the chromophore and the second order induced dipoles assuming the static atomic polarizability $\alpha\left(0\right)$. This contribution originates from interaction of the chromophore electronic state $a$ with states, where chromophore remains in the state $a$ but two environmental building blocks are excited:
\begin{gather} 
E_{ge,ge}^{pol2}\left(\alpha\left(0\right),\alpha\left(-\varepsilon_{ab}^{\left(A\right)}\right)\right) \approx 8\sum_{n,\eta}\sum_{n'\ne n,\eta'} \left\langle b,\eta_{n}\right|V\left|a,\boldsymbol{0}\right\rangle \nonumber \\
\times \frac{\left\langle a,\boldsymbol{0}\right|V\left|a,\eta_{n}\eta'_{n'}\right\rangle \left\langle a,\eta_{n}\eta'_{n'}\right|V\left|b,\eta_{n}\right\rangle }{\left(F_{0\eta}^{\left(n\right)}+F_{0\eta'}^{\left(n'\right)}\right)\left(F_{0\eta}^{\left(n\right)}+\varepsilon_{ab}^{\left(A\right)}\right)}.\label{eq:EnergyShift-unknown}
\end{gather}
Approximations required to obtain Eq. (\ref{eq:EnergyShift-SecondOrder})  are described in Supporting information (SI), together with the full formula for the second order correction to the transition energy shift.

\subsection{Transition dipole}

Transition dipole between the ground and the excited states of a chromophore embedded in its environment can be defined as $\boldsymbol{\mu}_{GE}=-e\left\langle G\right|\boldsymbol{r}\left|E\right\rangle$, where $\left|G\right\rangle$  and $\left|E\right\rangle$ are the ground and excited state of the whole system (chromophore+environment), and $\boldsymbol{r}$ are electronic coordinates of all electrons (chromophore+environment). The ground- and the excited state wavefunctions can be obtained by applying the same approach as for the transition energy shifts, i.e. perturbation expansion of the interaction operator. The first and the second order corrections to the transition dipole $\Delta\boldsymbol{\mu}_{GE}=\boldsymbol{\mu}_{GE}-\boldsymbol{\mu}_{ge}^{(A)}$, where $\boldsymbol{\mu}_{ge}^{(A)}$ is vacuum transition dipole, are obtained using Eq. (\ref{eq:DynamicPolar-def}) and the same approximations as for the excitation energy shift:
\begin{equation} \label{eq:Transition-dipole-env}
\begin{split}
\Delta\boldsymbol{\mu}_{GE}^{\left(1+2\right)}\approx\sum_{n}\boldsymbol{\mu}_{ge}^{\left(n\right)ind,1}\left(\overleftrightarrow{\alpha}_{dynamic}\left(\varepsilon_{ge}^{\left(A\right)}\right)\right) \\ +\sum_{n}\boldsymbol{\mu}_{ge}^{\left(n\right)ind,2}\left(\overleftrightarrow{\alpha}_{dynamic}\left(\varepsilon_{ge}^{\left(A\right)}\right)\right).
\end{split}
\end{equation}
Here, $\boldsymbol{\mu}_{ge}^{\left(n\right)ind,1}\left(\overleftrightarrow{\alpha}\right)$ is the dipole on environmental building block $n$ induced by transition charges (density) of the chromophore. The difference between the transition dipole of the chromophore embedded in environment and the chromophore in vacuum, corresponds to the dipole induced  on the environmental building blocks by the chromophore (vacuum) transition density, assuming dynamic polarizability at chromophore excitation energy. The full formula for the transition dipole change connected with the interaction with environment can be found in SI.

\subsection{Excitonic coupling}

The interaction energy between two chromophores embedded in an environment can be obtained from excited state energy splitting as $\left|J_{AB}\right|=\frac{1}{2}\sqrt{\left(E_{2}-E_{1}\right)^{2}-\left(\tilde{\varepsilon}_{B}-\tilde{\varepsilon}_{A}\right)^{2}}$, where $E_{i}$ is excited state energy of the whole interacting system, with $\tilde{\varepsilon}^{\left(A\right)}$ and $\tilde{\varepsilon}^{\left(B\right)}$ representing the excited state energies of individual chromophores embedded in the environment. The energies of the excited states are obtained using the same procedure as for the transition energy shift. In the case of two chromophore system, there is a possibility of degenerate (or close to degenerate) excited states of the chromophores, which would result in divergence of the perturbation expansion. The divergence can be avoided by application of the degenerate state perturbation expansion.  For a homodimer, or for chromophores with a very small difference in transition energies, the interaction energy $J_{AB}$ up to the second order of perturbation expansion, within the same approximations as in the case of the excitation energy shift calculation, reads as
\begin{equation} \label{eq: Interaction-energy-env}
\begin{split}
J_{AB}\approx J_{AB}^{vac}+E_{ge,ge}^{\left(AB\right),pol1}\left(\alpha_{dynamic}\left(\varepsilon_{ge}\right)\right) \\
+E_{ge,ge}^{\left(AB\right)pol,2}\left(\alpha_{dynamic}\left(\varepsilon_{ge}\right)\right) 
\\ +\frac{\left[E_{ge,gg}^{\left(AB\right)}-E_{ge,ee}^{\left(AB\right)}\right]\left[E_{ge,gg}^{\left(AB\right)}+E_{ge,ee}^{\left(AB\right)}+2E_{ge}^{\left(A\right)ele}\right]}{\varepsilon_{ge}}.
\end{split}
\end{equation}
Here, $J_{AB}^{vac}$ is the interaction energy between two chromophores in vacuum. The environmental contribution to the interaction energy corresponds to the interaction of the chromophore $A$ transition density with dipoles on environmental building blocks induced by the transition density of chromophore $B$ assuming dynamic atomic polarizabilities. The full formula for the homodimer interaction energy is shown in SI.

\subsection{Higher order terms}
Within the approximations used so far, adding higher order terms to perturbation expansion would result in adding higher order polarization terms $E_{ge,ge}^{\left(AB\right),pol,n}\left(\alpha_{dynamic}\left(E\right)\right)$ and $E_{ge,ge}^{\left(A\right),pol,n}\left(\alpha_{dynamic}\left(E\right)\right)$ to interaction energy and transition dipole, respectively. Summation of the whole perturbation expansion would result in a full polarization of the environment expressed through the dynamic atomic polarizabilities at the chromophore transition energies. For the excitation energy shift, one would need to account for all combinations of polarization of the environment expressed by both the frequency dependent and the static polarizabilities. For example, the third order correction to the excitation energy shift reads as
\begin{widetext}
\begin{equation} \label{eq:EnergyShift-ThirdOrder}
\begin{split}
&\Delta E_{g\rightarrow e}^{\left(3\right)}\approx\frac{1}{2}\left[E_{ee,ee}^{pol,3}\left(\alpha\left(0\right)\right)-E_{gg,gg}^{pol,3}\left(\alpha\left(0\right)\right)\right]+\left[E_{ee}^{env-pol,3}\left(\alpha\left(0\right)\right)-E_{gg}^{env-pol,3}\left(\alpha\left(0\right)\right)\right] \\
&+\frac{1}{16}\left[2E_{ge,ge}^{pol,3}\left(\alpha\left(\varepsilon_{ge}\right)\right)+3E_{ge,ge}^{pol2}\left(\alpha\left(0\right),\alpha\left(\varepsilon_{ge}^{\left(A\right)}\right),\alpha\left(\varepsilon_{ge}^{\left(A\right)}\right)\right)+3E_{ge,ge}^{pol2}\left(\alpha\left(0\right),\alpha\left(0\right),\alpha\left(\varepsilon_{ge}^{\left(A\right)}\right)\right)\right] \\
&-\frac{1}{16}\left[2E_{ge,ge}^{pol,3}\left(\alpha\left(-\varepsilon_{ge}\right)\right)+3E_{ge,ge}^{pol2}\left(\alpha\left(0\right),\alpha\left(-\varepsilon_{ge}^{\left(A\right)}\right),\alpha\left(-\varepsilon_{ge}^{\left(A\right)}\right)\right)+3E_{ge,ge}^{pol2}\left(\alpha\left(0\right),\alpha\left(0\right),\alpha\left(-\varepsilon_{ge}^{\left(A\right)}\right)\right)\right].
\end{split}
\end{equation}
\end{widetext}

\subsection{Classical polarizable atom model}
Approximating the environmental building blocks by quantum harmonic oscillators yields the same formulas for the environmental effects as in the general case of quantum polarizable atoms, however, with additional restrictions on the form of the frequency dependent polarizabilities. In harmonic oscillator approximation, the frequency dependent atomic polarizabilities fulfill the following relations: $\alpha\left(E\right)=\alpha\left(0\right)\left(1-E/\hbar\omega\right)^{-1}$ and $\alpha\left(-E\right)=\alpha\left(0\right)\left(1+E/\hbar\omega\right)^{-1}=\alpha\left(0\right)\left(2-\alpha\left(0\right)/\alpha\left(E\right)\right)^{-1}$, where $\omega$ is frequency of the harmonic oscillator. In the harmonic approximation, the system is therefore fully determined by its static and dynamic atomic polarizabilities. In quantum polarizable atom model, there is still a freedom in mutual relation between $\alpha\left(E\right)$ and $\alpha\left(-E\right)$, which are only coupled through $2\overleftrightarrow{\alpha}_{dynamic}\left(E\right)=\alpha\left(E\right)+\alpha\left(-E\right)$.

Classical polarizable atom model (as used for example in polarizable atom forcefield) is obtained from the perturbative treatment of the environment outlined above when we assume chromophore excitation energies large compared to the electrostatic interaction of the chromophore with its environment, but still small compared to the excitation energy of environment building blocks ($\varepsilon_{ge}\ll F_{0\eta}$). For that case we can neglect the frequency dependence of the polarizabilities $\left(F_{0\eta}^{\left(n\right)}-E\right)^{-1}\sim\left(F_{0\eta}^{\left(n\right)}\right)^{-1}$ and we obtain $\alpha\left(E\right)=\alpha\left(-E\right)=\alpha\left(0\right)$. Within this approximation, the excitation energy is shifted only due to a different electrostatic chromophore-environment interaction in the electronic ground and excited states, and due to different polarizations of the environment in the ground and excited states. The energy shift then reads as 
\begin{equation} \label{eq:EnergyShift-Classical}
\begin{split}
&\qquad \qquad \Delta E^{cl}_{g\rightarrow e}=\left(E_{ee}^{ele}-E_{gg}^{ele}\right) \\
&+\frac{1}{2}\left[E_{ee,ee}^{pol}\left(\overleftrightarrow{\alpha}\left(0\right)\right)-E_{gg,gg}^{pol}\left(\overleftrightarrow{\alpha}\left(0\right)\right)\right] \\
+&\left[E_{ee}^{env-pol}\left(\overleftrightarrow{\alpha}\left(0\right)\right)-E_{gg}^{env-pol}\left(\overleftrightarrow{\alpha}\left(0\right)\right)\right].
\end{split}
\end{equation}
The excitation energy shift is no longer dependent on the chromophore transition density. The transition dipole change due to the interaction with the environment corresponds to a dipole induced by chromophore transition density in the environment $\Delta\vec{\mu}_{GE}=\sum_{n}\vec{\mu}_{ge}^{\left(n\right)ind}\left(\overleftrightarrow{\alpha}\left(0\right)\right)$. The excitonic interaction energy between two chromophores embedded in classical polarizable atom environment consist of vacuum interaction energy plus interaction with dipoles in the environment induced by the transition density of the other chromophore expressed only through the static atomic polarizabilities $J_{AB}=J_{AB}^{vac}+E_{ge,ge}^{\left(AB\right),pol}\left(\overleftrightarrow{\alpha}\left(0\right)\right)$.

\section{Results and discussion}
\label{sec:discussion}
In Section \ref{sec:theory} we have introduced the quantum polarizable atom model. In this section, we use it to obtain excited state properties of impurities in FG. 

\begin{figure}[tbp]
  \centering
  \includegraphics[width=0.5\textwidth]{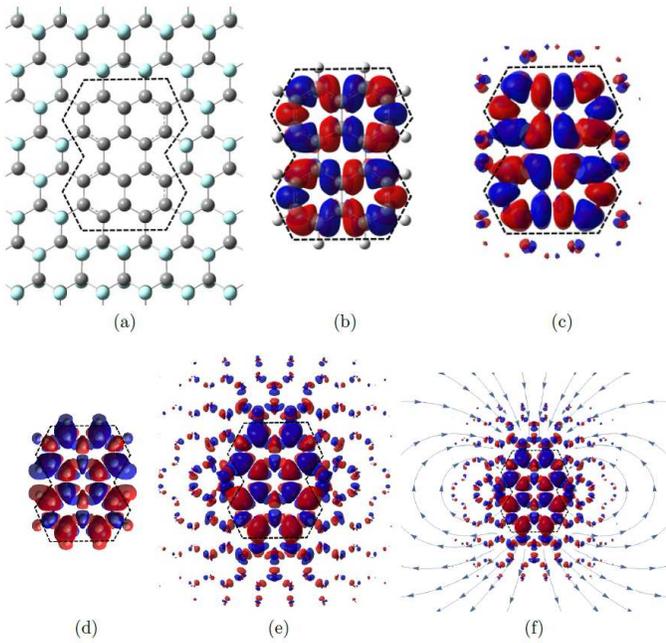}
  \caption{Comparison of transition densities of perylene molecule in vacuum and perylene-like impurity in FG. (a) Perylene defect in FG, (b) Perylene $S_{0}\rightarrow S_{2}$ transition in vacuum, (c) Perylene-like impurity in FG $S_{0}\rightarrow S_{2}$ transition (without atoms), (d) Perylene $S_{0}\rightarrow S_{1}$ transition in vacuum, (e) Perylene-like impurity in FG $S_{0}\rightarrow S_{1}$ transition (without atoms), (f) Perylene impurity transition density together with electric field from transition density of isolated molecule (without atoms).}
\label{fig:FG-Perylen-TrDens}
\end{figure}

An inspection of the molecular orbitals of disordered FG reveals confinement of $\pi$-molecular orbitals in the area of the impurity (see Fig. \ref{fig:FG-Perylen-TrDens}). The reason for this confinement is the separation of $\pi$-conjugated structure of the impurity and $\sigma$-bond character of the pure FG \cite{Slama2018}. The transition densities, the highest occupied and the lowest unoccupied electronic orbitals for "defective" FG are well localized in the area of the impurity, and they are similar to corresponding quantities from isolated impurity-like molecule. For the calculation of optical properties, this makes the impurities treatable as individual molecules embedded in pure FG environment (Fig. \ref{fig:FG-Perylen-TrDens}). The leakage of the transition density from the impurity into the FG is proportional to the corresponding transition dipole. For forbidden transitions the transition density is localized only on the defect whereas for transitions with large dipole moment, small part of the transition density is delocalized over few rows of surrounding FG atoms. This leakage can be interpreted as dipoles induced in FG by the impurity transition density. This conclusion is also supported by the shape of the transition density in FG area, which follows the transition dipole electric field (Fig. \ref{fig:FG-Perylen-TrDens}f). This correspondence allows us to treat the impurity as a molecule, and describe the effects of the FG environment on its excited state properties effects using the quantum polarizable atom model developed in this paper.
%
%
The quantum polarizable atom model was applied here to calculate excitation energy and transition dipole shifts together with interaction energies for "graphen-like" impurities on FG surface. The results obtained by the quantum polarizable atom model are compared with full QC calculation of excited state properties, Poisson-TrEsp method \cite{Renger2012}, harmonic oscillator model and classical polarizable atom model on FG clusters with different impurity sizes, shapes and mutual orientation. For testing we have used a set of FG clusters with perylene-, anthanthrene-, bisanthrene- and peropyrene-like impurities. The corresponding isolated molecules are the smallest symmetric aromatic hydrocarbons which have the excitation energies smaller than FG energy gap and the lowest electronic excited state allowed for the optical excitation. These properties make the tested moleculoids  good candidates for centers for absorption and excitation energy transfer in FG artificial light harvesting antenna \cite{Slama2018}. For the calculation of interaction energies between excited states of individual impurities, the symmetric clusters with two identical impurities in different distances and mutual orientations were used. The total testing set consisted of 31 different clusters with size ranging from 400-800 atoms.


\begin{figure}[tbp]
  \centering
  \includegraphics[width=0.5\textwidth]{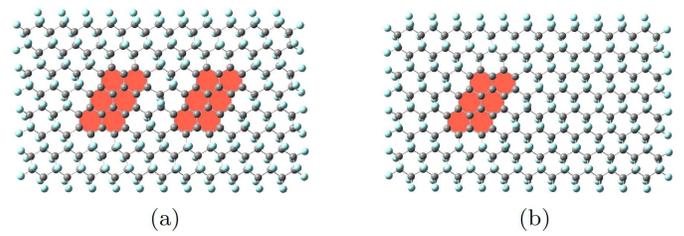}
\caption{Example structures of the cluster with anthanthrene impurity. (a) Cluster with two impurities for interaction energy calculation. (b) Cluster with single impurity used for excitation energy and transition dipole shift calculations.}
\label{fig:Cluster-example}
\end{figure}

\begin{figure}[tbp]
  \centering
  \includegraphics[width=0.48\textwidth]{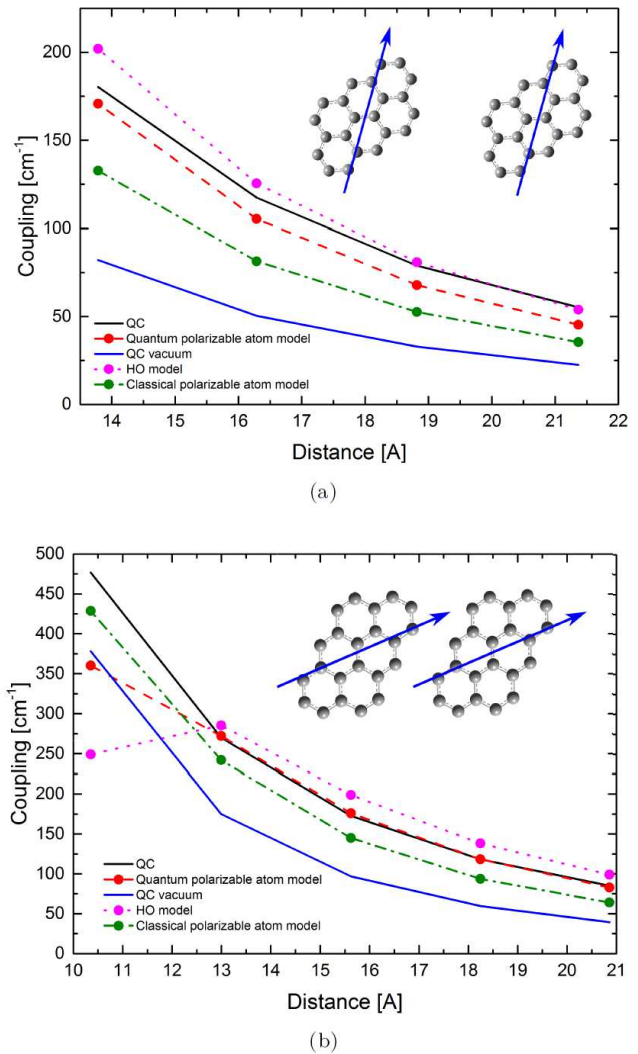}	
\caption{Comparison of different approximate approaches for excitonic coupling calculation with QC results on FG sheet with two anthanthrene impurities with (a) parallel and (b) serial (head-to-tail) orientation of transition dipoles.}
\label{fig:Coupling-anth}
\end{figure}

Structures with molecule-like defects were created from finite FG sheets in an ideal periodic geometry (C-C distance 1.594A) \cite{Zboril2010}, leaving always at least three rows of fluorinated graphene between the moleculoids and the edges of the FG sheet to minimize the effects of its finite size. Because there is no clear way how reliably account for the FG environment effects using standard methods, e.g. transition density cube method, interaction energies between impurities were extracted from QC calculation of excited state energy splitting. According to Frenkel exciton model \cite{van2000photosynthetic}, the interaction energy for symmetric homodimer is equal to half of the excited state splitting. To avoid numerical errors, the calculation of interaction energies from excited state splitting requires the use of symmetry constrains for structure optimization of the whole cluster to obtain identical effects of the FG environment for both impurities. Excitation energies and transition dipole shifts induced by the fluorographene environment were obtained as differences between excitation energies and transition dipoles calculated for FG clusters containing single impurities and the corresponding isolated impurity-like molecules in vacuum. For these calculations we have used the geometries and positions of the impurities from the dimer structures (Fig. \ref{fig:Cluster-example}), where always one impurity was replaced by pure FG. Geometry of this structure was then optimized while keeping the geometry of the other impurity fixed. This allows us to directly compare the results for systems with single and two impurities, and to investigate the effects of the the presence of an other impurity on the impurity excited state properties. Quantum chemistry calculations were performed using Gaussian 09 package \cite{gaussian}. The geometry of the FG clusters with impurities were optimized using DFT approach with B3LYP functional and LANL2DZ valence basis set. During the geometry optimization the border carbons of a cluster were kept frozen at positions corresponding to the pure FG periodic structure, to simulate the effect of an infinite fluorographene sheet, in which large structure deformations due to the presence of an impurity would not be possible. Excited state energies and transition dipoles were obtained using TD-DFT approach with $\omega$B97XD long range corrected hybrid functional and LANL2DZ basis set. The combination of DFT methods for geometry optimization and excited state calculation was taken from our previous study \cite{Slama2018}, where we have shown its good performance for excited states of isolated hydrocarbons. Interaction energies for different mutual configurations of impurities on FG together with interaction energies of corresponding molecules in vacuum at the same geometry and position as in FG were calculated. Representative example of excitonic coupling dependence on mutual distance between impurities is shown in Fig. (\ref{fig:Coupling-anth}) for two mutual orientations of athanthrene impurities. The results for other impurities can be found in SI. From these results it can be seen, that the interaction energies between impurities in FG are enhanced compared to vacuum interaction energies. This enhancement is even larger for larger distances between the impurities. This behavior, which is opposite to what is observed for chromophores embedded in protein environments, can be attributed to the 2D structure of the FG sheet (see discussion in SI). The same is true also for transition dipole moment, which is enhanced in FG compared to the vacuum value.
%
  
%

The ground- and the excited state charges and the transition atomic charges for the quantum polarizable atom model were obtained from RESP \cite{RESP} fitting of the ground-, excited- and transition state electrostatic potentials of isolated molecules in vacuum at optimized vacuum geometries, using only carbon atoms for the fitting. The FG ground state charges were obtained from RESP fitting of the ground state potential of a pure FG cluster in the direction perpendicular to FG surface for three positions around the center of the cluster (in the center of the ring unit, above fluorine atom and above carbon atom in opposite direction to the C-F bond). Cutoff distance of 3\AA from the nearest atom center  was used for electrostatic potential fitting, yielding the inner carbon charge $q_{C}^{in}$=-0.0522 and an opposite fluorine charge ($q_{F}$=$-q_{C}$). For the border carbon atoms (with two fluorine atoms attached) a twice larger charge than for inner carbons was assumed to compensate for the charge of additional fluorine atom. The details of the fitting procedure can be found in SI, where the effects of the charge on results obtained by quantum polarizable atom model are also discussed.

\begin{figure}[tbp]
  \includegraphics[width=0.48\textwidth]{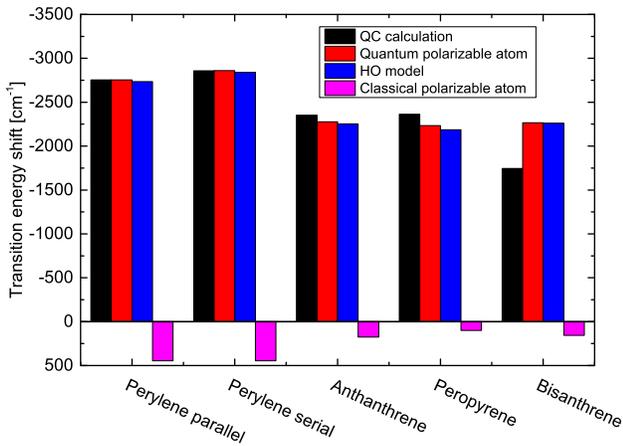}
  \caption{Transition energy difference ($\Delta E_{g\rightarrow e}$) between impurity in FG and impurity-like molecule in vacuum calculated by different methods.}
  \label{fig:Eshift}
  
\end{figure}
\begin{figure}[tbp]
  \includegraphics[width=0.48\textwidth]{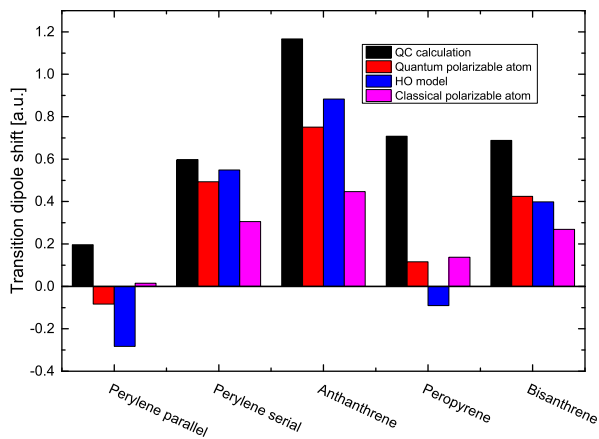}
  \caption{Transition dipole difference (${\mu}_{GE}-{\mu}_{ge}$) between impurity in FG (${\mu}_{GE}$) and impurity-like molecule in vacuum (${\mu}_{ge}$) calculated by different methods.}
  \label{fig:TrDip}
\end{figure}

For impurities we tested, the excitation energies were found to be much higher then their variation between individual impurities. We can therefore simplify the model and use the same polarizabilities $\alpha\left(E\right)$ for all impurities. In order to keep the system as simple as possible, we treat the fluorographene C-F group as a single coarse-grained polarizable atom at a position of the carbon atom. Using this simplified model, the static atomic polarizability is obtained from fitting static polarizabilities of small pure FG clusters of different sizes and shapes, calculated using DFT aproach with $\omega$B97XD functional and LANL2DZ basis set. Details of the fitting procedure can be found in SI. Assuming an isotropic polarizability in FG plane, the remaining four free parameters of the model, parallel and perpendicular polarizability $\alpha\left(E\right)$ and $\alpha(-E)$ to the FG surface, are obtained from fitting the QC interaction energies and excitation energy shifts. The resulting coarse-grained atomic polarizabilities are shown in Table \ref{tbl:Atomic-polarizabilities}, and the corresponding interaction energies, excitation energy and transition dipole shifts are shown in Figs. \ref{fig:Coupling-anth}-\ref{fig:TrDip}. Complete results can be found in SI.

\begin{figure}[tbp]
  \includegraphics[width=0.48\textwidth]{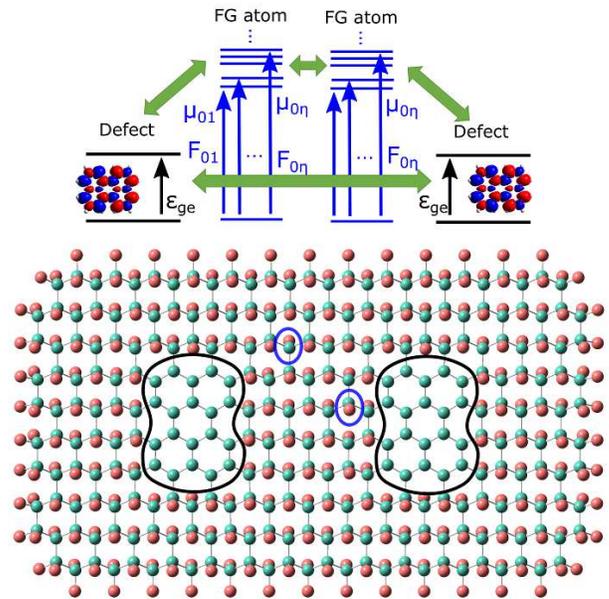}
\caption{Graphical representation of the quantum polarizable atom model as used on disordered FG.}
\label{fig:GraphPolarAtModel}
\end{figure}

\begin{table}
  \caption{Atomic polarizabilities for C-F coarse-grained atoms in atomic units}
  \label{tbl:Atomic-polarizabilities}
  \centering
  \begin{tabular}{lll}
    \hline
    			& in FG plane & perpendicular to FG \\
    \hline
    $\alpha\left(0\right)$   & 5.10   & 5.17\\
    $\alpha\left(E\right)$   & 15.49  & 6.09\\
    $\alpha\left(-E\right)$  & 0.56  & 1.70\\
    $\alpha_{HO}\left(E\right)$ & 17.43 & 5.19\\
    $\alpha_{HO}\left(-E\right)$ \quad & 2.99 & 5.14\\
    \hline
  \end{tabular}
\end{table}
The excitation energy shifts for perylene-like impurities were calculated for two different shapes of the FG sheet around the impurity. The structure denoted as parallel is made of FG sheet elongated in the direction perpendicular to perylene impurity transition dipole, whereas for the structure denoted as serial is elongated in direction of the transition dipole. Correct description of the excitation energy difference between these two cases confirms the validity of our treatment of environment effects.


For calculation with the quantum polarizable atom model, we have used only the second order perturbation expansion, because the terms which were neglected during the derivation of Eqs. (\ref{eq:EnergyShift-SecondOrder}), (\ref{eq:Transition-dipole-env}) and (\ref{eq: Interaction-energy-env}) have a magnitude similar to the higher order polarization terms $E_{ab}^{pol}\left(\alpha\left(E\right)\right)$. Including the third order polarization terms $E_{ab}^{pol3}\left(\alpha\left(E\right)\right)$ results in only a small change of the excited state properties of impurity embedded in FG environment, suggesting that second order perturbation expansion of the system-environment interaction is sufficient for FG systems.

Interaction energies from the quantum polarizable atom model are compared with results obtained with the harmonic oscillator approximation, the classical approximation ($\alpha\left(E\right)=\alpha\left(0\right)$) and the Poisson-TrEsp method \cite{Renger2012}. In Poisson-TrEsp method, the FG sheet is described as a dielectric layer with impurities in vacuum cavities. The interaction energies between the excited states of the impurities are calculated by numerical solution of Poisson equation. Position of the impurities and atomic charges were used the same as for the quantum polarizable atom model calculation, and the thickness and relative permitivity of the dielectric slab were obtained from the best fit of the QC results. From the comparison of classical polarizable atom model with results from QC calculation, it can be seen that although the classical model (MMpol) accounts for environmental effects on interaction energies and transition dipoles reasonably well (Figure \ref{fig:Coupling-anth} and \ref{fig:TrDip}), it provides wrong shifts of transition energies (Figure \ref{fig:Eshift}). For FG systems, the best agreement with full QC calculation is obtained for quantum polarizable atom model. The description of the environment using frequency dependent polarizabilities is therefore crucial to obtain the right description of the environment effects on interaction energies, transition dipoles and excitation energy shifts for fluorographene systems. 

The quantum polarizable atom model is a perturbative and therefore inherently approximative approach open to improvements. The highest discrepancy between quantum polarizable atom model and full QC calculation in this work was observed for interaction energies at the smallest distances between impurities. Using non-isotropic polarizabilities in FG plane had only little effect on the results, non-isotropic polarizabilities cannot therefore explain the observed discrepancy. One possible reason for the discrepancy could be our treatment of the impurities as static charge distributions which cannot be polarized by other impurities. This in turn  results from our treatment of impurities as two level systems. Including higher excited states of the impurities into the perturbation expansion would lead to the "polarization" of the impurity. A better approach, however, would be to include electric field of induced dipoles in the environment into the QC calculation of impurity excited state properties. Mutual polarization of the system and the environment is then resolved using self consistent procedure as in polarizable QM/MM scheme. Other improvement might be enabled by application of a full atomic model instead of a coarse-grained description of C-F atoms as a single polarization center. Because of the symmetry of the system it would, however, require more sophisticated fitting scheme to obtain unique static polarizabilities for carbon and fluorine atoms.

\section{Conclusions}
\label{sec:conclusions}
In this paper, we derived the quantum polarizable atom model of the environment using perturbation treatment of the system-environment coupling.
Within this approach we described the environment as a set of polarizable atoms with frequency dependent polarizablities. The interaction energies, site excitation energy and transition dipole shifts are obtained using zero order ground, excited and transition atomic charges from quantum chemical calculation of chromophores in vacuum. Assumptions needed to obtain this simple model are quite general and satisfied by many artificial as well as biological systems. Standard polarizable QM/MM approach for treating the environmental effects on chromophore excited state properties can be obtained from the presented approach by assuming the static atomic polarizabilities instead of frequency dependent ones. We applied the quantum polarizable atom model to calculate interaction energies, site energy shift and transition dipole shifts for impurities on fluorographene obtaining a good quantitative agreement with full quantum chemistry calculation. Comparison of the quantum chemistry results with the classical polarizable atom model reveals the need to use frequency dependent polarizabilities to properly account for environment induced excitation energy shifts in fluorographene based systems.

Quantum polarizable atom model provides a generalization of the polarizable QM/MM approach and the next logical step in a more precise description of the environment effects. This approach might be particularly useful for systems where standard approaches fails to explain the observed experimental behavior, especially for the systems where the excitation energy of the chromophore is not negligible compared to the excitation energy of environment building blocks. The present approach is in general not limited to the second order approximation, and it offers obvious pathways towards its extension and improvements.

\begin{acknowledgements}
This work was supported by the Czech Science Foundation (GA\v{C}R) grant no. 18-18022S and by the Neuron Fund for Support of Science, grant Neuron Impuls for Physics 2014. V.S. acknowledges support by the Grant Agency of Charles University (GAUK) grant no. 1162216.
\end{acknowledgements}


%

\end{document}